\documentclass[journal]{vgtc}

\onlineid{0}
\vgtccategory{Research}
\vgtcpapertype{Technique}

\title{The Plot Thickens: Quantitative Part-by-Part Exploration\\ of MLLM Visualization Literacy}

\author{
    \authororcid{Matheus Valentim}{0000-0003-0860-2084},
    \authororcid{Vaishali Dhanoa}{0000-0002-0493-8616},
    \authororcid{Gabriela Molina Le{\'o}n}{0000-0002-9223-2022}, and
    \authororcid{Niklas Elmqvist}{0000-0001-5805-5301}
}

\authorfooter{
\item Matheus Valentim, Vaishali Dhanoa, Gabriela Molina Le{\'o}n, and Niklas Elmqvist are with Aarhus University in Aarhus, Denmark. E-mail: \{au763015, dhanoa, leon, elm\}@cs.au.dk. 
}

\shortauthortitle{The Plot Thickens}

\abstract{%
    Multimodal Large Language Models (MLLMs) can interpret data visualizations, but what makes a visualization understandable to these models? Do factors like color, shape, and text influence legibility, and how does this compare to human perception?
    In this paper, we build on prior work to systematically assess which visualization characteristics impact MLLM interpretability.
    We expanded the Visualization Literacy Assessment Test (VLAT) test set from 12 to 380 visualizations by varying plot types, colors, and titles.
    This allowed us to statistically analyze how these features affect model performance.
    Our findings suggest that while color palettes have no significant impact on accuracy, plot types and the type of title significantly affect MLLM performance.
    We observe similar trends for model omissions. 
    Based on these insights, we look into which plot types are beneficial for MLLMs in different tasks and propose visualization design principles that enhance MLLM readability.
    Additionally, we make the extended VLAT test set, VLAT\textbf{\textit{ex}}, publicly available on \url{https://osf.io/ermwx/} together with our supplemental material for future model testing and evaluation.
}


\teaser{
    \centering
    \includegraphics[width=\linewidth]{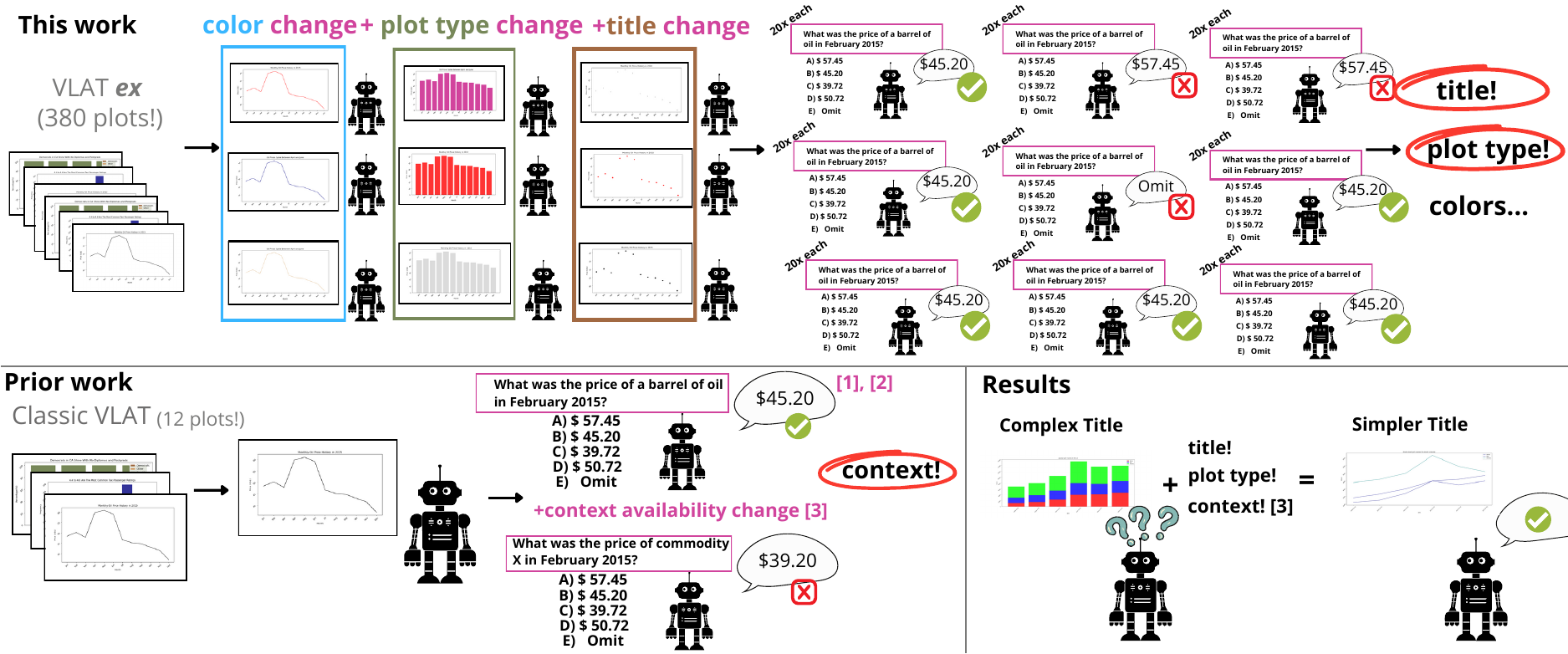}
    \caption{\textbf{Visualization literacy for MLLMs.}
    Our work advances the field's knowledge about \textit{which} charts and chart components play a role in the visualization literacy of multimodal large language models.
    These findings suggest a set of common design principles that can help practitioners optimize data visualizations specifically for MLLMs.
    }
    \label{fig:teaser}
}

\graphicspath{{figures/}{pictures/}{images/}{./}}

\usepackage[svgnames, dvipsnames, table]{xcolor}

\usepackage{enumitem}

\usepackage{svg}

\usepackage{fontawesome5}

\usepackage{booktabs}

\usepackage{listings}

\usepackage{colortbl}

\usepackage{float}

\usepackage{amsmath}

\usepackage[most]{tcolorbox}
\usepackage{xspace}

\usepackage[table]{xcolor} 
\usepackage{array} 
\usepackage{booktabs} 
\usepackage{textcomp}

\usepackage{enumitem}   

\usepackage{soul}
\definecolor{new-green}{rgb}{0.104,0.667,0.229}
\newcommand{\del}[1]{} 

\newcommand{\chip}[2]{%
  \tcbox[
    enhanced,
    frame empty,
    interior style={fill=#1!20},
    colback=#1!20,
    colframe=#1,
    sharp corners=northeast,
    sharp corners=southwest,
    arc=2mm,
    boxrule=0.0pt,
    left=0.5mm,
    right=0.5mm,
    top=-0.1mm,
    bottom=-0.1mm,
    fontupper=\footnotesize\sffamily,
    nobeforeafter,
    tcbox raise base,
    valign=center,
    size=minimal,
    height=0.8\baselineskip
  ]{#2}%
}

\begin{document}

\maketitle

\section{Introduction}

The rise of multimodal large language models (MLLMs) with their capacity to ``see''---or rather, translate images into vector embeddings---opens up a world of possibility for the use of these models for visualization.
This new capability has prompted researchers to develop benchmarks that measure how well MLLMs interpret visualizations, drawing on established visualization literacy assessment methods for humans~\cite{DBLP:journals/tvcg/BoyRBF14, lee_vlat_2017, borner_data_2019}.
Accordingly, several recent studies have presented early results on the visualization literacy of LLMs, including using different visualization literacy tests~\cite{DBLP:journals/tvcg/BendeckS25}, comparing the performance of different models~\cite{Li2024}, varying the chart types involved~\cite{hong25llm}, and studying  MLLM performance for misleading charts~\cite{chen2025unmaskingdeceptivevisualsbenchmarking, Tonglet2025misleading, pandey2025benchmarkingvisuallanguagemodels}.

Among these, Li et al.~\cite{Li2024}, Bendeck and Stasko~\cite{DBLP:journals/tvcg/BendeckS25}, and Hong et al.~\cite{hong25llm}  provide important insights into MLLM performance and highlight how certain plot types and question formats affect their capabilities.
They raise critical questions about the influence of external information, misleading elements, and crucial to this work, they begin to discuss counterfactuals of why models fail to comprehend certain visualizations, raising visualization colors and geometry as candidates.

However, by relying on mostly limited samples and contraposition, these works lack systematic investigation into why MLLMs perform well on certain aspects of visualization literacy but poorly on others, and what interventions might improve their capabilities.
We address this knowledge gap by conducting an in-depth experiment as follows:

\begin{itemize}
    \item\textbf{Two foundation models:} We involve two separate MLLMs: Google Gemini Pro 2.0 and GPT-4o; 
    
    \item\textbf{Generalized charts:} Starting from the classic VLAT~\cite{lee_vlat_2017}, we generalize the questions to use all 12 chart types and color codings in the test, yielding 1,380 cases rather than the original 60; 

    \item\textbf{Reliability and pass@k:} We perform 20 repetitions and use the pass@k metric---a measurement of how often the model gets a correct answer within $k$ attempts---as the reliability metric; and
    
    \item\textbf{General and Localized Effects:} We run regressions to derive broad statistical insights, while also conducting subgroup regressions to improve control and understand one-to-one changes.

\end{itemize}

We discover significant evidence that plot types and the title's wording affect MLLMs' performance, while the color palette does not. We also establish that for specific visualization tasks, certain plot types are significantly more effective than others.
Drawing on these findings, we then discuss their likely causes, and derive unified visualization design principles that are common to both humans and MLLMs on the ranking of visual channels, the optimal use of color palettes, and the impact of titles and legends on chart comprehension. 

In summary, our work contributes to the field of data visualization and MLLMs by:
(1) Presenting robust statistical estimators quantifying whether some of the main visual elements on a plot (such as plot type and color palette) have an impact on MLLMs' visualization understanding. 
(2) Shedding statistically sound light on which plot type alterations should be made in specific visualization tasks to improve MLLMs' literacy.
(3) providing an extended VLAT dataset, and the code on how to make it, in order to help future work better evaluate the visualization literacy of MLLMs on specific visual elements that make up a plot.
\section{Related Work}

Here we review research on visualization literacy, plot readability, and methods for assessing visual understanding in humans and models.

\subsection{Visualization Literacy and Assessment}
\label{subsec:vislit}

\textit{Visualization literacy}~\cite{DBLP:journals/tvcg/BoyRBF14}---the ability to read, interpret, and derive meaning from visual data representations---has become increasingly vital in our information-rich society~\cite{borner_data_2019}.
As data visualizations proliferate across news media, scientific publications, and workplace settings, the capacity to understand and extract insights from these representations has evolved from a specialized skill to a fundamental competency.

Different researchers conceptualize visualization literacy with nuanced emphases while converging on core principles.
Most definitions center on extracting meaningful information from visualizations, with some highlighting the resistance to misleading visual elements and others encompassing the ability to not only interpret but also construct visualizations~\cite{ge23}.
This multifaceted understanding of visualization literacy complements the concept of visualization readability, which refers to the inherent properties of a visualization that make it more or less comprehensible.
While literacy refers to the viewer's capability, \textit{readability} describes the visualization's accessibility~\cite{cabouat_previs_2024}.

The theoretical foundations of visualization readability draw from the Grammar of Graphics framework~\cite{Wilkinson2005}, which decomposes visualizations into fundamental visual elements whose interplay creates meaningful data representations.
As Bertin's seminal work established, these elements---including titles, axes, color palettes, and plot types---combine with external context to guide viewers toward data understanding.
Empirical studies demonstrate how manipulating these elements significantly impacts perception and interpretation; for example, that excessive color brightness can impair comprehension, or that visualization titles strongly influence viewers' takeaways~\cite{kong18}.

Given the importance of visualization literacy, researchers have developed various assessment frameworks to evaluate this skill.
Among these, the Visual Literacy Assessment Test (VLAT)~\cite{lee_vlat_2017} has emerged as a particularly valuable resource.
VLAT consists of 12 visualizations with 53 questions spanning diverse assessment tasks---from identifying extrema to recognizing patterns. 
VLAT's comprehensive coverage of visualization types and question formats, coupled with its open availability, has made it a benchmark for visualization literacy research.

\paragraph{Our contribution.}

We build directly on the VLAT framework by expanding it from 12 to 380 visualizations with systematic variations in plot types, colors, and titles.
This expanded dataset creates a more comprehensive testing ground for MLLMs, enabling statistical analysis of which visualization characteristics most impact model performance, which was not previously possible with existing frameworks.

\subsection{Factors Affecting Visualization Interpretation}
\label{subsec:factors}

Understanding how viewers interpret visualizations requires examining the interplay of perceptual, cognitive, and design factors that influence this process.
It may also offer ideas into how human perception differs from that of AI models.
The systematic study of graphical perception emerged through empirical work by Eells et al.~\cite{Eells1926} and Croxton et al.~\cite{Croxton1927, Croxton1932}, who investigated how people viewed statistical graphics and compared plot types.
These early investigations culminated in Cleveland and McGill's seminal work~\cite{cleveland_graphical_1984, cleveland_experiment_1986, cleveland_graphical_1987}, which systematically ranked visual encoding channels based on their effectiveness for reading values.
Mackinlay extended this work to develop automated chart construction systems~\cite{mackinlay_automating_1986}, while later research by Stewart and Best~\cite{stewart_examination_2010} consolidated the original ten rankings into four more general categories.
Modern eye-tracking technologies have further enhanced our understanding of human perception by revealing what causes confusion in charts~\cite{lalle16confusion}, which visual patterns benefit recall~\cite{borkin16recall}, how visual saliency serves as a measure of attention~\cite{riche13metricssaliency}, and methods for assessing visualization proficiency~\cite{toker14untrainedvisusers}.
These studies provide empirical evidence for design principles that enhance visualization effectiveness.

Pinker's theory of graph comprehension~\cite{pinker_theory_1990} proposes that graphs are processed hierarchically through schemas of structures, encodings, and messages, and that unfamiliar schemas require more cognitive resources than familiar patterns.
Carpenter and Shah~\cite{carpenter_model_1998} proposed a sequential process model where graph comprehension emerges from pattern recognition to meaning construction.
Halford et al.~\cite{halford_how_2005} established that humans can process only four variables simultaneously without performance degradation, providing an important cognitive constraint for visualization design.
Taking a more practical approach to the topic, Shin et al.~\cite{shin23scannerdeeply} built deep learning models from crowdsourced eye-tracking data to simulate human gaze patterns on visualizations.

While visual elements form the foundation of charts, textual components---particularly titles---significantly impact interpretation.
Eye-tracking studies by Borkin et al.~\cite{borkin_what_2013} revealed that viewers spend more time on text elements, especially titles, than on other visualization components.
Kong et al.~\cite{kong18} showed that visualization titles heavily influence viewers' takeaways from charts.
This aligns with Hullman and Diakopoulos' work on visualization rhetoric~\cite{DBLP:journals/tvcg/HullmanD11}, which highlighted how textual annotations guide attention and shape narrative framing.
These findings connect to broader cognitive biases in information processing, including selective perception~\cite{dearborn_selective_1958}, confirmation bias~\cite{nickerson_confirmation_1998}, and biased assimilation~\cite{lord_biased_1979}, which affect not only which aspects of charts viewers attend to but also how they interpret the presented data.

\paragraph{Our contribution.}

Our work systematically examines how perceptual and cognitive factors translate to MLLM visualization comprehension.
By studying how specific visual properties (plot types, color schemes, titles) affect model performance, we provide insights into whether the same factors that influence human interpretation also impact machine understanding.
This comparison reveals both similarities and differences between human and MLLM perception, highlighting visualization design principles that can benefit both humans and MLLMs.

\subsection{LLMs and Multimodal Capabilities}
\label{subsec:llms-multimodal}

The emergence of Large Language Models (LLMs) marks a significant advancement in artificial intelligence.
Research demonstrates that scale---in terms of model parameters, training data, and computational resources---serves as the primary driver of performance improvements~\cite{brown2020language, wei2022emergent, ganguli2022predictability}.
Models with billions or even trillions of parameters, such as GPT-4~\cite{OpenAI2023GPT4TR}, PaLM~\cite{chowdhery2022palm}, LLaMA~\cite{touvron2023llama}, and Gemini~\cite{Anil2023gemini}, have demonstrated remarkable capabilities across diverse tasks.

The standard development paradigm for LLMs involves pre-training on vast text corpora using self-supervised learning objectives like next-token prediction, followed by additional training phases to align the models with human preferences.
Pre-training typically occurs at organizations with substantial computational resources, with models then deployed either as services (like ChatGPT) or open-source offerings (like Vicuna~\cite{chiang2023vicuna}, LLaMA~\cite{touvron2023llama}, or DeepSeek R1).

A pivotal evolution in language model architecture has been the combination of multiple modalities, particularly vision and language, creating Multimodal Large Language Models (MLLMs).
These models can process and reason about both textual and visual inputs simultaneously.
Early multimodal systems employed separate encoders for different modalities with limited integration, but recent architectures like GPT-4~\cite{OpenAI2023GPT4TR} and Gemini~\cite{Anil2023gemini} incorporate deeper cross-modal connections, enabling more sophisticated visual reasoning.
However, the process by which MLLMs interpret visual data differs fundamentally from human perception.
While humans leverage specialized visual processing systems developed through evolution and experience, MLLMs convert visual inputs into token embeddings within the same representational space as text.
Haehn et al.~\cite{haehn_evaluating_2019} examined how convolutional neural networks (CNNs) perform on graphical perception tasks, finding that while CNNs can sometimes match human performance, they are not good models for human graphical perception.

\paragraph{Our contribution.}

Our work systematically isolates specific visual properties---shapes, color palettes, and contextual elements such as titles and legends---to determine which factors most significantly impact MLLM visual understanding.
This approach reveals both similarities and differences between human and machine perception of image data, contributing to a more nuanced understanding of how MLLMs process visual information and informing the development of principles that work effectively for both human and machine interpreters.

\subsection{Chart Question Answering}

Chart Question Answering (CQA) takes a natural language question along with the chart as input and provides a natural language answer as output. 
Hoque et al.~\cite{hoquesurvey} surveyed the literature on CQA, discussing its different types of input and output dimensions, such as factual vs.\ open-ended textual queries resulting in fixed vs.\ open vocabulary answers, single vs.\ multiple views for visualization based queries resulting in textual answers, and multimodal input resulting in multimodal output.

Previous work, such as FigureQA~\cite{kahou2017figureqa}, uses synthetic images for five main plot types to ask questions and receive answers in a fixed "yes" or "no" vocabulary. 
Meanwhile, DVQA~\cite{kafle2018dvqa} provides a dataset and an algorithm that helps in answering open-ended questions related to bar charts, better than existing visual question answering algorithms~\cite{elzer2011automated, malinowski2014multi}. 
OpenCQA~\cite{kantharaj2022opencqa} takes any chart type and an open-ended question as input to provide open vocabulary answers using extractive and generative models to enhance chart interpretation. 
Kim et al.~\cite{kimexploringchart} also use open-ended queries, but focusing on assisting blind and low vision (BLV) users in understanding visualizations.
Wu et al.~\cite{wuchartinsights} compile a large-scale dataset for low-level CQA tasks (e.g., characterizing distributions, finding extremum) and assess the performance of both open source (e.g., LLaVA~\cite{liu2023visual}) and closed source MLLMs (e.g., Qwen-VL-Plus~\cite{Bai2023QwenVLAF}, GPT-4-vision preview~\cite{OpenAI2023GPT4TR}) on these tasks. 
Zeng et al.~\cite{zengCQA} also use open and closed sources MLLMs on existing CQA tasks to understand the challenges that MLLMs face while solving complex reasoning visualization tasks. 
Most recent and quite closely related to our work, are the works by Bendeck and Stasko~\cite{DBLP:journals/tvcg/BendeckS25}, Li et al.~\cite{Li2024}, and Hong et al.~\cite{hong25llm} which use VLAT to examine the visualization literacy of LLMs using fixed vocabulary; more on these below. 

\paragraph{Our contribution.}
 
Our work provides factual text-based queries (extended VLAT dataset) along with single view visualizations as input to MLLMs which results in fixed vocabulary answers.
We focus on the visualization literacy of MLLMs using the new dataset.

\subsection{Benchmarking MLLM Visualization Literacy}
\label{sec:llmliteracy}

Benchmarking approaches for MLLMs typically fall into two categories: ground-truth evaluations that measure performance against predetermined correct answers, and LLM-as-judge frameworks where models evaluate responses to open-ended questions.
Early work in MLLM visual capabilities focused on fundamental perceptual skills rather than visualization literacy specifically.
These studies examined how models perceive geometric shapes ~\cite{Guo2024UnderstandingGP} and colors ~\cite{HyeonWoo2024VLMsEE} as building blocks.

Several pioneering studies have directly assessed visualization literacy in MLLMs.
Bendeck and Stasko~\cite{DBLP:journals/tvcg/BendeckS25} evaluated multiple MLLMs using the VLAT dataset, analyzing performance across different question types and identifying areas where models excel or struggle.
Li et al.~\cite{Li2024} extended this work by investigating failure modes and analyzing what factors might cause MLLMs to misinterpret visualizations.
Both studies provided insights into MLLM visualization literacy but relied on limited samples and lacked systematic variation of visual elements.

Hong et al.~\cite{hong25llm} significantly advanced the field by implementing a more rigorous methodology.
Their work employed repeated testing---having models respond to each question hundreds of times to establish reliability metrics.
While this approach improved statistical reliability, it still used the original 12 VLAT visualizations without systematically varying visual elements like plot types, colors, and contextual cues.
Their primary contribution was creating alternative versions of VLAT without real-world references to investigate whether models were using pre-trained knowledge rather than actually interpreting visualizations.

Taking a different angle, Tonglet et al.~\cite{Tonglet2025misleading} examined MLLM vulnerability to misleading visualizations---charts that distort underlying data through techniques such as truncated or inverted axes.
Their findings revealed that such distortions dramatically reduced question-answering accuracy to random-baseline levels.
This work highlights the importance of understanding how visualization design choices impact MLLM performance and suggests that models may rely heavily on visual conventions rather than extracting underlying data relationships.
Their solution---extracting data tables and using text-only LLMs for interpretation---further indicates that current MLLMs struggle with certain aspects of visual parsing that humans readily overcome.

In very recent work, Chen et al.~\cite{chen2025unmaskingdeceptivevisualsbenchmarking} introduced the Misleading ChartQA Benchmark, a dataset with 3,000+ examples across 21 ``misleader types'' and 10 chart types to evaluate MLLM abilities to detect and interpret misleading visualizations.
They rigorously benchmarked 16 state-of-the-art MLLMs against their dataset, revealing significant limitations in identifying visually deceptive practices.
While they focus specifically on deceptive visualization detection and correction, we take a broader approach by examining how visualization characteristics (plot types, colors, titles) fundamentally affect MLLM comprehension across both standard and potentially misleading contexts.

Another recent work by Pandey and Ottley~\cite{pandey2025benchmarkingvisuallanguagemodels} explores how MLLMs interpret visualizations through systematic benchmarking of four models (GPT-4, Claude, Gemini, and LLaMa) using VLAT~\cite{lee_vlat_2017} and CALVI~\cite{ge23}.
Their findings reveal that while models show competence in basic chart interpretation, all struggle with identifying misleading visualization elements.
However, unlike our work, they do not study the impact of chart type and elements on performance.

Despite these contributions, current benchmarking approaches exhibit several limitations.
First, most studies examine a limited number of visualization variations for the same data, making it difficult to isolate specific visual features that influence model performance.
While Hong et al. tested models hundreds of times on the same visualizations, they did not systematically vary visual properties across a large set of visualization instances.
Second, few studies control for the impact of contextual elements such as titles, labels, and color schemes in a statistically rigorous manner.
Third, existing studies primarily focus on overall accuracy rather than analyzing patterns of errors and omissions that might reveal deeper insights into how MLLMs process visualizations.

\paragraph{Our contribution.}

Our work addresses these methodological gaps by:
(1) expanding the test from 12 to 380 visualizations with controlled variations in chart types, colors, and titles;
(2) implementing rigorous reliability assessment through 20 repetitions and pass@k metrics; and
(3) isolating specific visual elements to determine their individual effects on MLLM performance.
Unlike previous work that repeated tests on a small set of visualizations, our approach systematically varies visual properties across a much larger visualization corpus, enabling analysis of which design factors impact MLLM comprehension.
\section{Method}
\label{sec:method}

\newcommand{\linechart}{\chip{Orange}{\faChartLine~Line chart}\xspace}
\newcommand{\barchart}{\chip{SteelBlue}{\faChartBar~Bar chart}\xspace}
\newcommand{\piechart}{\chip{DeepPink}{\faChartPie~Pie chart}\xspace}
\newcommand{\scatterplot}{\chip{Chartreuse}{\faBraille~Scatterplot}\xspace}
\newcommand{\stackedbarchart}{\chip{Cerulean}{\faLayerGroup~Stacked bar chart}\xspace}
\newcommand{\percbarchart}{\chip{Gold}{\faPercent~Percentage bar chart}\xspace}
\newcommand{\histogram}{\chip{Olive}{\faBars~Histogram}\xspace}

\newcommand{\colorpalette}{\chip{Purple}{\faPalette~Color palette}\xspace}
\newcommand{\titlefactor}{\chip{Brown}{\faHighlighter~Title}\xspace}

Our experiment aims to understand the impact of visualization characteristics on Multimodal Large Language Models' (MLLMs) ability to interpret charts.
To systematically analyze this relationship, we expanded the Visualization Literacy Assessment Test (VLAT) dataset through controlled variations of visual elements such as chart types, color palettes, and titles.
We then evaluated two state-of-the-art MLLMs---Google Gemini 1.5 Flash and GPT-4o---on this expanded dataset using multiple-choice questions.
Below we present the details.
The dataset, protocol, and full results for this experiment can be found on OSF: \url{https://osf.io/ermwx/}

\subsection{Expanding the VLAT Dataset}
\label{sec:expandingVLATdataset}

The standard Visualization Literacy Assessment Test~\cite{lee_vlat_2017} consists of 53 questions on 12 visualizations, covering a range of chart types such as bar charts, line charts, and pie charts.
The questions assess five distinct analytical skills: retrieving values, finding extrema, comparing values, determining ranges, and finding correlations/trends.
Each question is multiple choice with four answer options plus an \textit{omit} option.

To systematically evaluate how visual characteristics affect MLLM performance, we expanded the original VLAT dataset. 
Creating a completely new dataset would give us a more adequate structure, but designing a balanced set of questions to assess distinct visual literacy skills is complex and risky.
Evaluations require careful testing to avoid overly easy questions, ambiguous categorization, or misalignment with visualization literacy goals.
Additionally, shifting to a new dataset could break continuity with prior visualization literacy research, making it harder to compare results and build on recent findings.
Thus, rather than creating an entirely new VLAT dataset specialized for MLLM benchmarking, we felt there was value in retaining the existing dataset. 

To expand the dataset, we created variations along three dimensions:

\begin{itemize}[noitemsep, topsep=0pt, parsep=0pt, partopsep=0pt]

    \item\textbf{Chart types:} We used the underlying data from each original visualization to generate alternative chart representations where dimensionally possible (Table~\ref{tab:plot-types}).
    
    \item\textbf{Color palettes:} We implemented ten different color schemes ranging from vibrant to muted.
    Color palettes are sequence of colors within a same pattern (e.g., grayish colors, with different tones of gray, black and light yellow, or neon colors, with bright neon-like green, purple and other highly saturated colors).
    
    \item\textbf{Titles:} We created both neutral and suggestive title variations for each visualization. Suggestive titles hint at some characteristic of the visualization (e.g. ``Oil Prices Spike Between April and June'' instead of a regular title ``Monthly Oil Price History in 2015'' or ``Samsung Leads, Apple Second in Global Phone Market Share'' instead of ``Global Smartphone Market Share (\%)'').

\end{itemize}

\begin{figure}[htb]
    \centering
    \begin{subfigure}[b]{0.5\columnwidth}
        \centering
        \includegraphics[width=\textwidth]{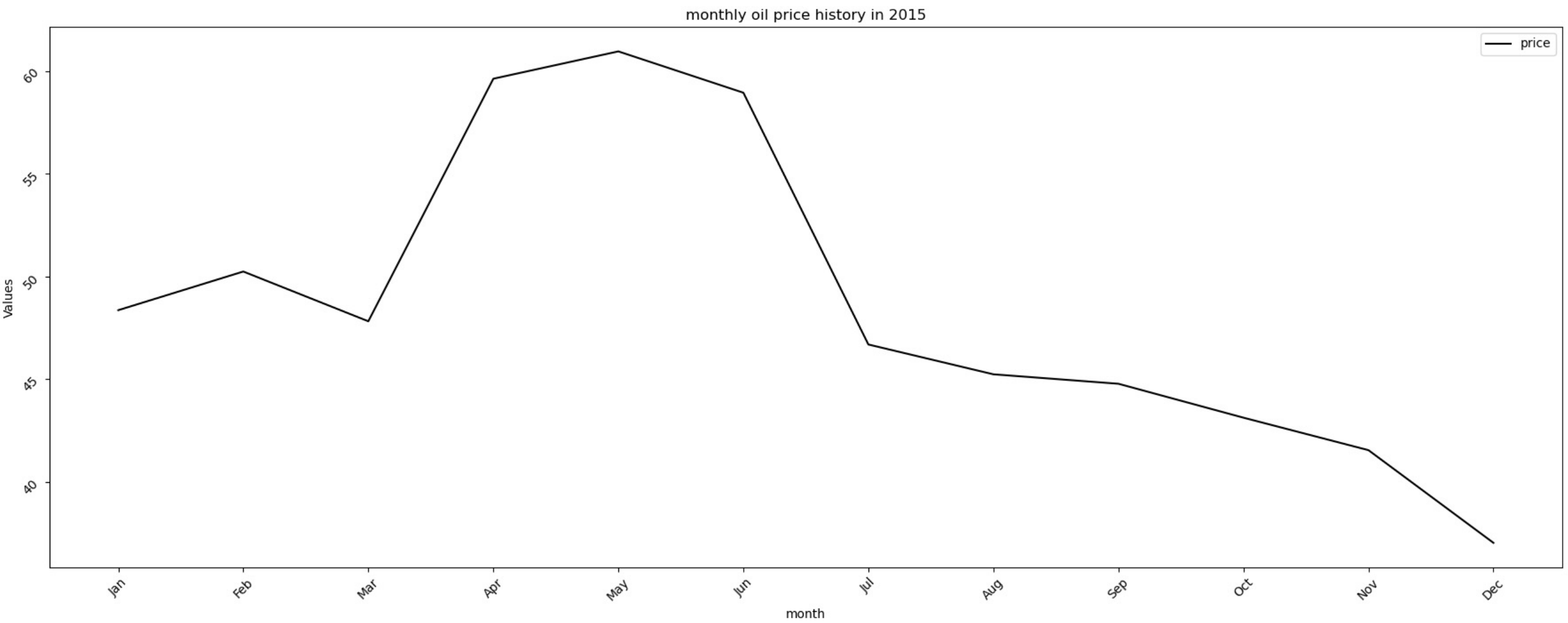}
        \caption{\textbf{Original VLAT line plot.}
        }
        \label{fig:original-line-plot}
    \end{subfigure}%
    \begin{subfigure}[b]{0.5\columnwidth}
        \centering
        \includegraphics[width=\textwidth]{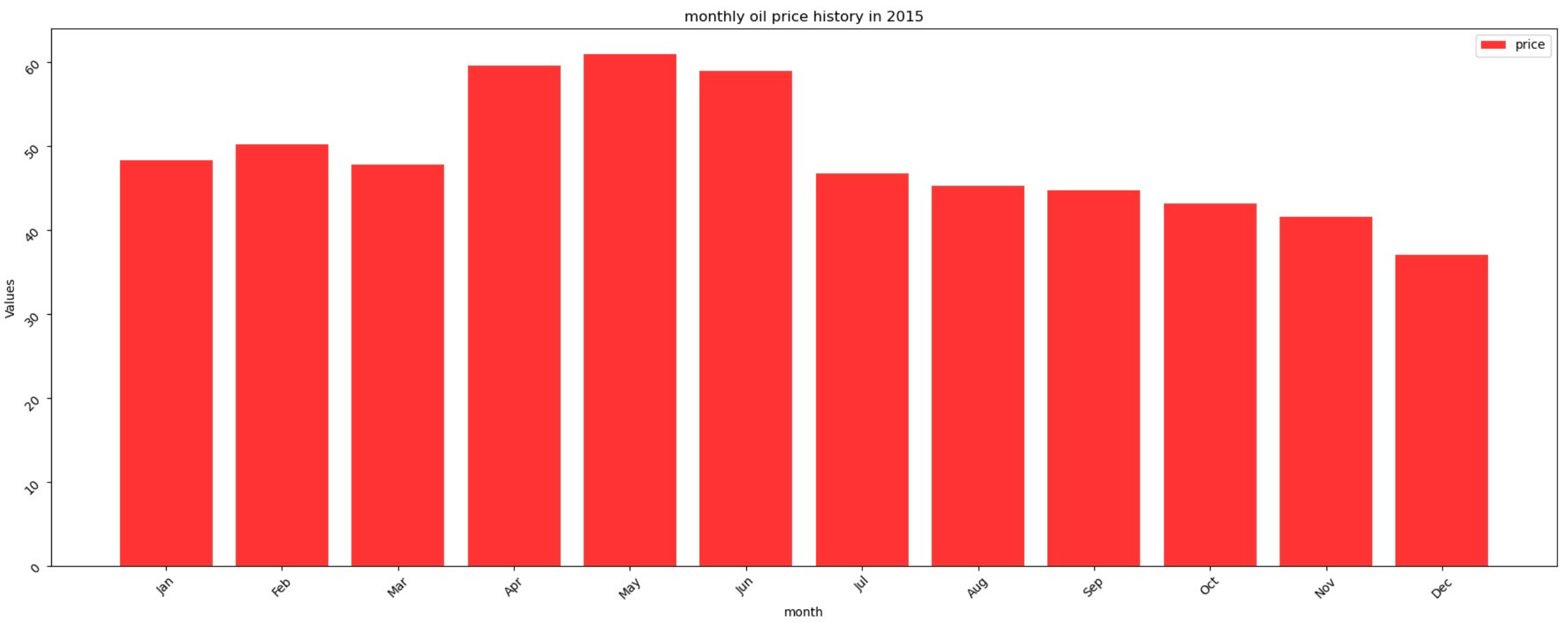}
        \caption{\textbf{Altered line plot.}
        Bar plot, different color palette.}
        \label{fig:altered-bar-plot}
    \end{subfigure}%
    \subfigsCaption{\textbf{Examples of original and altered plots.}
        VLAT plot and altered plot, both visualizing monthly oil price.}
    \label{fig:original-altered}
\end{figure}

\Cref{fig:original-line-plot} shows an original VLAT line chart, while \cref{fig:altered-bar-plot} demonstrates a derived variation using a bar chart representation with a different color palette.
The derivation of new chart types was constrained by the dimensional properties of the original data.
For example, a line chart showing a time series could be transformed into a bar chart or scatterplot, but not into a stacked bar chart, which would require an additional categorical dimension.
\Cref{tab:plot-types} summarizes the allowable chart type transformations for each original visualization type.

\begin{figure}[htb]
    \centering
    \includegraphics[width=\linewidth]{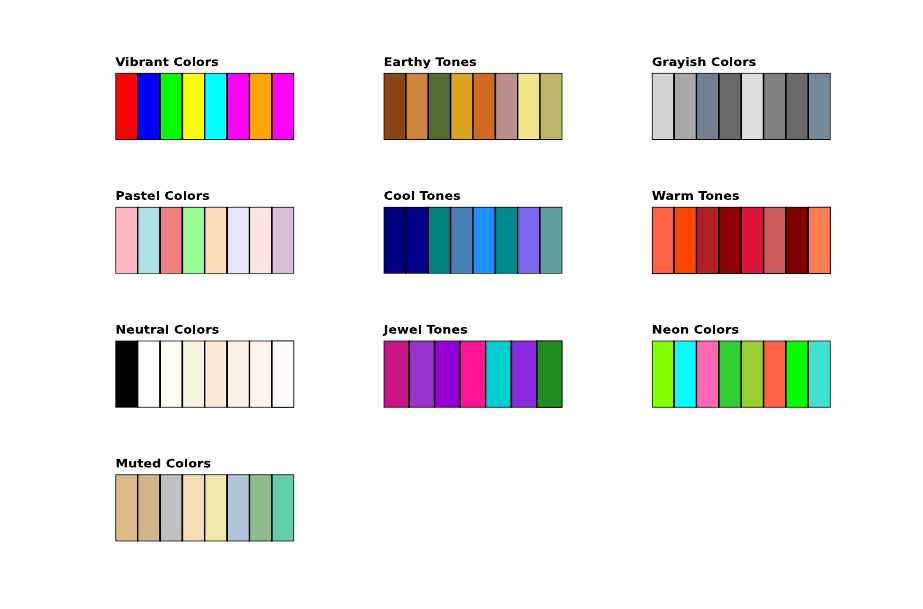} 
    \caption{\textbf{Color palettes.}
    The ten color palettes used in our experiment.
    }
    \label{fig:color-palettes}
\end{figure}

Unlike chart type transformations, our color palette and title variations had no dimensional constraints.
For each chart, we created:
(1) Ten different color palette variations (as detailed in \cref{fig:color-palettes});
(2) Two title versions (\textit{neutral} and \textit{suggestive}), and as many plot type variations as possible, as detailed in Table~\ref{tab:plot-types}.
Each chart maintained all the questions that its underlying dataset had in the initial VLAT, guaranteeing that question-plot pairing was the same as in the original dataset. 

\begin{table}[tbh]
    \centering
    \caption{\textbf{Plot variations.} Allowed plot types for each original plot type.}
   \arrayrulecolor{lightgray}
    \begin{tabular}{m{3cm}p{45mm}}
        \toprule
        \textbf{Original Plot Type} & \textbf{Allowed Plot Types} \\ 
        \midrule
        \histogram & \histogram \\ \hline
        \scatterplot & \scatterplot \\ \hline
        \percbarchart & \percbarchart \\ \hline
        \barchart & \barchart \scatterplot \\  \hline
        \piechart & \barchart \scatterplot \newline \piechart \\ \hline
        \linechart & \barchart \scatterplot\newline  \linechart \\ \hline
        \stackedbarchart & \stackedbarchart \linechart \newline \percbarchart  \\ \hline
        
        \bottomrule
    \end{tabular}
    \label{tab:plot-types}
\end{table}

Through this systematic expansion, we grew the original 12 VLAT visualizations into \textsc{VLAT ex}, a comprehensive test set of 380 visualizations and 3,220 rows, each representing a unique plot-question combination.
Three visualizations were excluded from this process.
These---a treemap, a United States Map and a Bubble Chart---were not considered interchangeable or comparable enough to the other visualizations.
Additionally, four questions were excluded due to an author mistake.
This expansion enabled us to evaluate MLLM performance across a controlled space of visualization variations while preserving the original VLAT's underlying data and question types.

\subsection{Testing the MLLMs on the Expanded Dataset}
\label{sec:testingLLMs}

We evaluated two state-of-the-art MLLMs---Google Gemini 1.5 Flash and GPT-4o---on our expanded VLAT \textit{ex} dataset. 
While our approach follows previous studies~\cite{hong25llm, Li2024, DBLP:journals/tvcg/BendeckS25} in using multiple-choice questions, our work differs significantly in scale.
With 380 visualizations (compared to the original 12), our evaluation provides a substantially larger foundation for analysis of visualization literacy patterns (Fig.~\ref{fig:teaser}).

To minimize confounding effects from prompt engineering, we maintained minimal and consistent instructions across all tests, simply asking the models to \textit{``answer the following question.''}
We also included the \textit{Omit} option to allow the models to abstain from answering when uncertain, enabling us to measure both accuracy and omission behavior.

Recognizing the stochastic nature of MLLM outputs, we repeated each visualization-question pair 20 times to obtain robust performance estimates.
This yielded a dataset where each row represents a unique visualization-question combination with the following attributes:

\begin{itemize}[noitemsep, topsep=0pt, parsep=0pt, partopsep=0pt]
    \item \textbf{Visualization-specific:} Chart type, color palette, title type;
    \item \textbf{Question information:} Question text, type, skill category;
    \item \textbf{Model performance:} Accuracy, omission rate; and
    \item \textbf{Model identifier:} Gemini 1.5 Flash or GPT-4o.
\end{itemize}

This dataset structure enabled us to perform detailed statistical analyses on how different visualization attributes affect MLLM performance across various analytical tasks.
By conducting 20 trials per visualization-question pair, we mitigated the effects of model randomness and established more reliable performance metrics for our analysis.

\subsection{Building the Different OLS Models}
\label{sec:probit}

To quantify how visualization attributes affect MLLM performance, we constructed several Ordinary Least Squares (OLS) regression models.
OLS regression provides interpretable coefficients that estimate each attribute's effect on performance while holding other attributes constant.
We built two sets of models with different dependent variables:

\begin{itemize}[noitemsep, topsep=0pt, parsep=0pt, partopsep=0pt]
    \item \textbf{Accuracy:} Using normalized counts of correct answers; and
    \item \textbf{Omission:} Using normalized counts of omitted answers.
\end{itemize}

We chose not to use more complex accuracy scores, i.e., using the difficulty rating provided by VLAT itself, as those were based on human performance, which doesn't necessarily align with what is hard for MLLMs.
Our general model specification takes the form:

\begin{equation}
Y^* = X\beta + \varepsilon, \quad \varepsilon \sim \mathcal{N}(0,1).
\end{equation}

where $Y^*$ represents the normalized dependent variable (either accuracy or omission rate), $X$ is the matrix of explanatory variables, $\beta$ is the vector of coefficients to be estimated, and $\varepsilon$ is the error term.

Our vector of covariates, X, are variables that qualify the specific combination of question and visualization in a given row.
These include visualization qualifications such as type of plot (e.g. bar plot, stacked bar plot), type of title (e.g., suggestive title), and color palette (e.g. grayscale); question qualifications, such as type of question (e.g., ``Find a maximum,'' ``Observe an underlying pattern'') and other controls such as underlying VLAT dataset and MMLM model (e.g. GPT-4o, Google Gemini).
In order to look for other possible transmission channels of our covariates' impact, we often included interactions between some of those variables (e.g., ``Retrieve Value'' Type of Question and ``Stacked Bar Plot'').
We chose columns interactions based on statistics of the best and worst performing task of each plot type according to simple descriptive statistics (e.g., \linechart and comparisons questions and \linechart and determine range questions).
We also sometimes included interactions in which a given plot type was believed to be more (or less) suited for, according to the original VLAT paper (e.g., the promising \stackedbarchart and value retrieval questions and the unfavored \histogram and value retrieval).

All of the covariates described above are binary variables; that is, they assume values of either 0 or 1.
To avoid multicolinearity we thus leave out one of the binaries for each category.

In the end, our model looks like this: 
\begin{equation}
\begin{aligned}
    Y^* &= \beta_0 + \beta_1 \text{(Chart Type)} + \beta_2 \text{(Title Type)} + \beta_3 \text{(Color Palette)} \\
        &\quad + \beta_4 \text{(Question Type)} + \beta_5 \text{(Dataset)} + \beta_6 \text{(MLLM Model)} \\
        &\quad + \beta_7 \text{(Interactions)} + \varepsilon, \quad \varepsilon \sim \mathcal{N}(0,1).
\end{aligned}
\end{equation}

We conducted our analysis at two levels and using two different techniques: \textbf{Full dataset analysis}, using all visualization-question pairs to identify broad patterns and \textbf{effects-coded regressions}; and \textbf{Chart-specific analysis}, creating separate models for subsets of the data containing specific chart types, using \textbf{dummy-coded regressions}.
The full dataset analysis allows to derive sound statistical meaning on the effect of the different variables in a larger sample with a larger variety of plots.
We rely on effects-coding so that our coefficients are not interpreted based on a specific plot type, but differences based on the models' grand mean---the mean of all observations in the dataset.   
The chart-specific analysis allows us to further investigate which plot type transformations are beneficial for the MLLMs, as in each subset there were only plot types that could be converted into one another.
In this approach, we implement a dummy-coded regression to get specific insight into the effect of each plot type in regards to the others.

This two-tier analysis offers both broad insights across all visualization types and detailed findings within specific chart categories.
It also mitigates a key limitation: due to dimensional constraints, not all plots can be transformed into every type.
As a result, a classic dummy-coded regression with one plot type omitted would produce misleading coefficients, since the excluded chart might not have been a viable transformation for the one under investigation. 

\section{Results}

Here we present our findings on how different visualization characteristics affect MLLM performance.
We organize our results into the impact of plot type, the influence of \colorpalette, the effect of \titlefactor, and performance differences when substituting one plot type for another.
For each area, we analyze multiple findings regarding accuracy (i.e., the number of correct answers) and omission rates (i.e., how often the models chose not to answer) across our VLAT \textit{ex} dataset.

\subsection{Impact of Plot Type}

To begin our presentation on the impact of plot type on MLLM performance, we first analyzed accuracy and omission rates across all questions grouped by plot type.
For each question, we tracked (1) the number of correct answers and (2) the number of times the models chose to omit an answer across that questions' 20 repetitions, creating performance distributions for each plot type.

\begin{figure*}[htb]
    \centering
    \begin{subfigure}{0.48\textwidth}
        \centering
        \includegraphics[width=\textwidth]{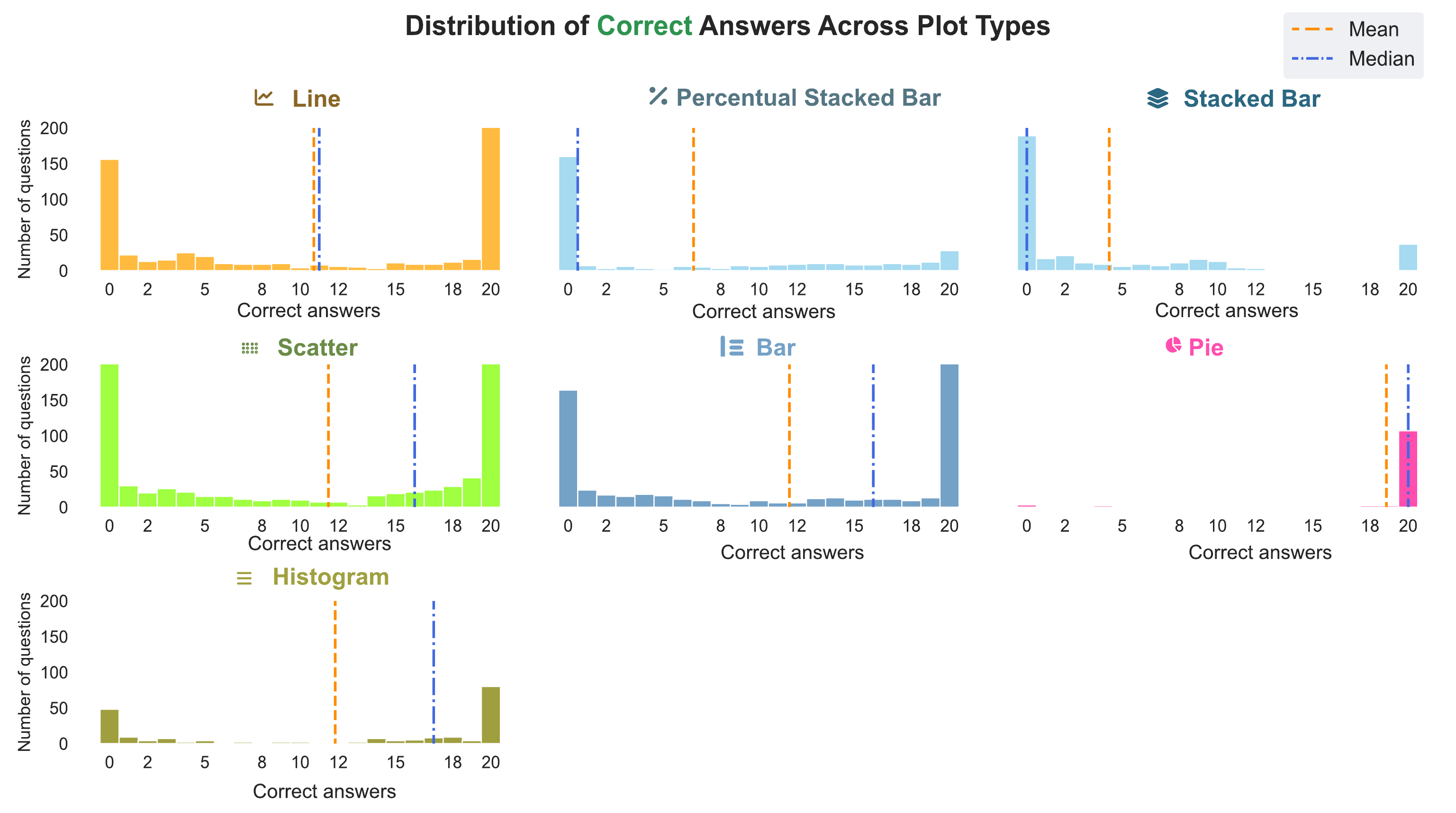}
        \caption{\textbf{Accuracy by plot type.}
        }
        \label{fig:correct-plot}
    \end{subfigure}
    \hfill
    \begin{subfigure}{0.48\textwidth}
        \centering
        \includegraphics[width=\textwidth]{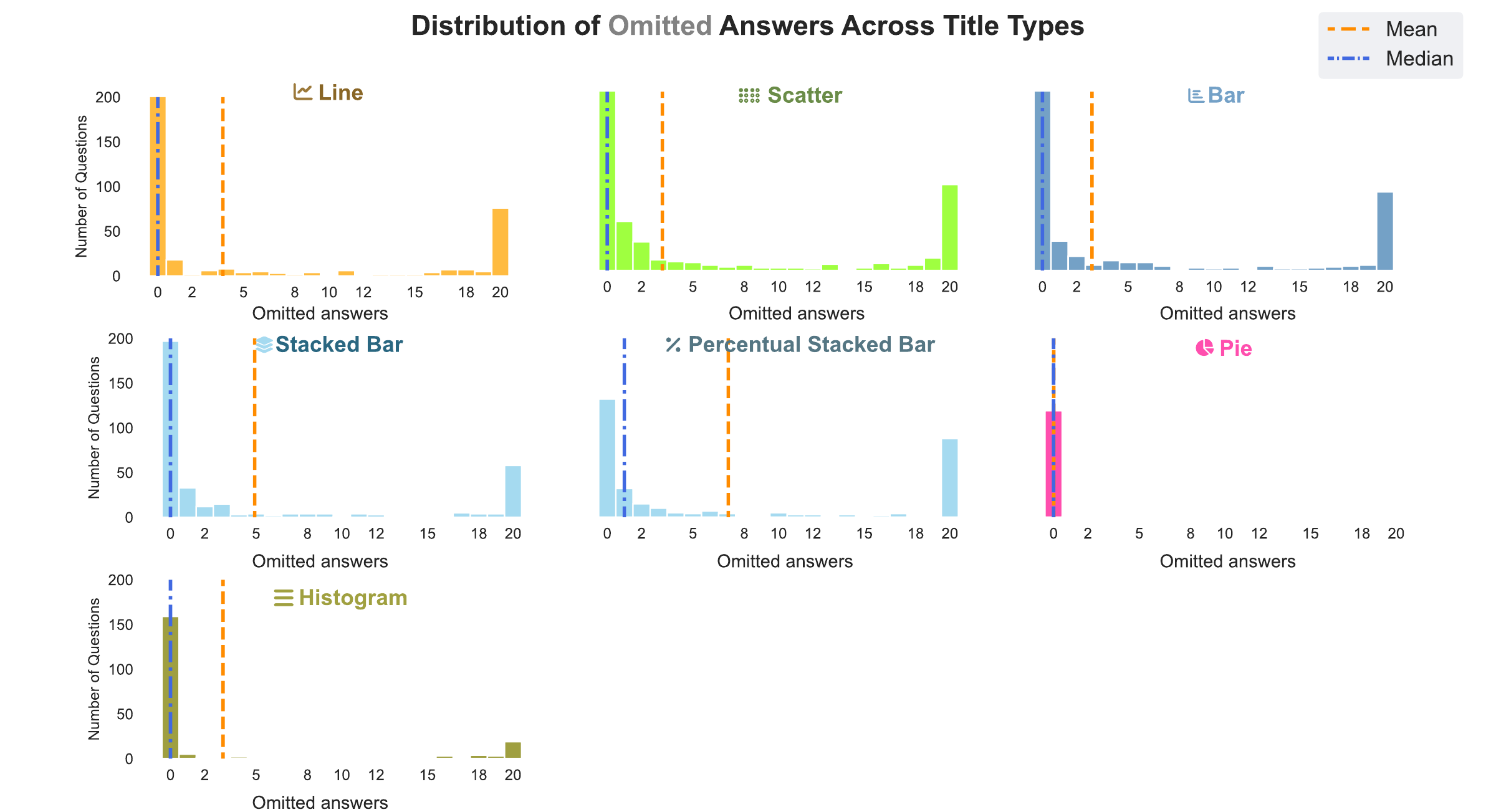}
        \caption{\textbf{Omissions by plot type.}
        }
        \label{fig:omission-plot}
    \end{subfigure}
    \caption{\textbf{Correct and omitted answers by plot type.}
    Accuracy and omission counts distribution per plot type.
    }
    \label{fig:acc_plt_type}
\end{figure*}

\Cref{fig:acc_plt_type} reveals distinct performance patterns across visualization types.
\linechart, \scatterplot, and \barchart showed similar performance distributions, with questions averaging between 10.6 and 16 correct responses out of 20 attempts.
\piechart demonstrated superior performance, with a higher proportion of questions receiving 20 correct attempts and minimal omissions.
In contrast, \stackedbarchart and \percbarchart yielded the poorest performance in both accuracy (means of 6.1 and 6.9 correct answers, respectively) and omission rates (means of 5 and 7 omitted answers, respectively). 

\begin{figure}[htb] 
    \centering
    \includegraphics[width=0.5\textwidth]{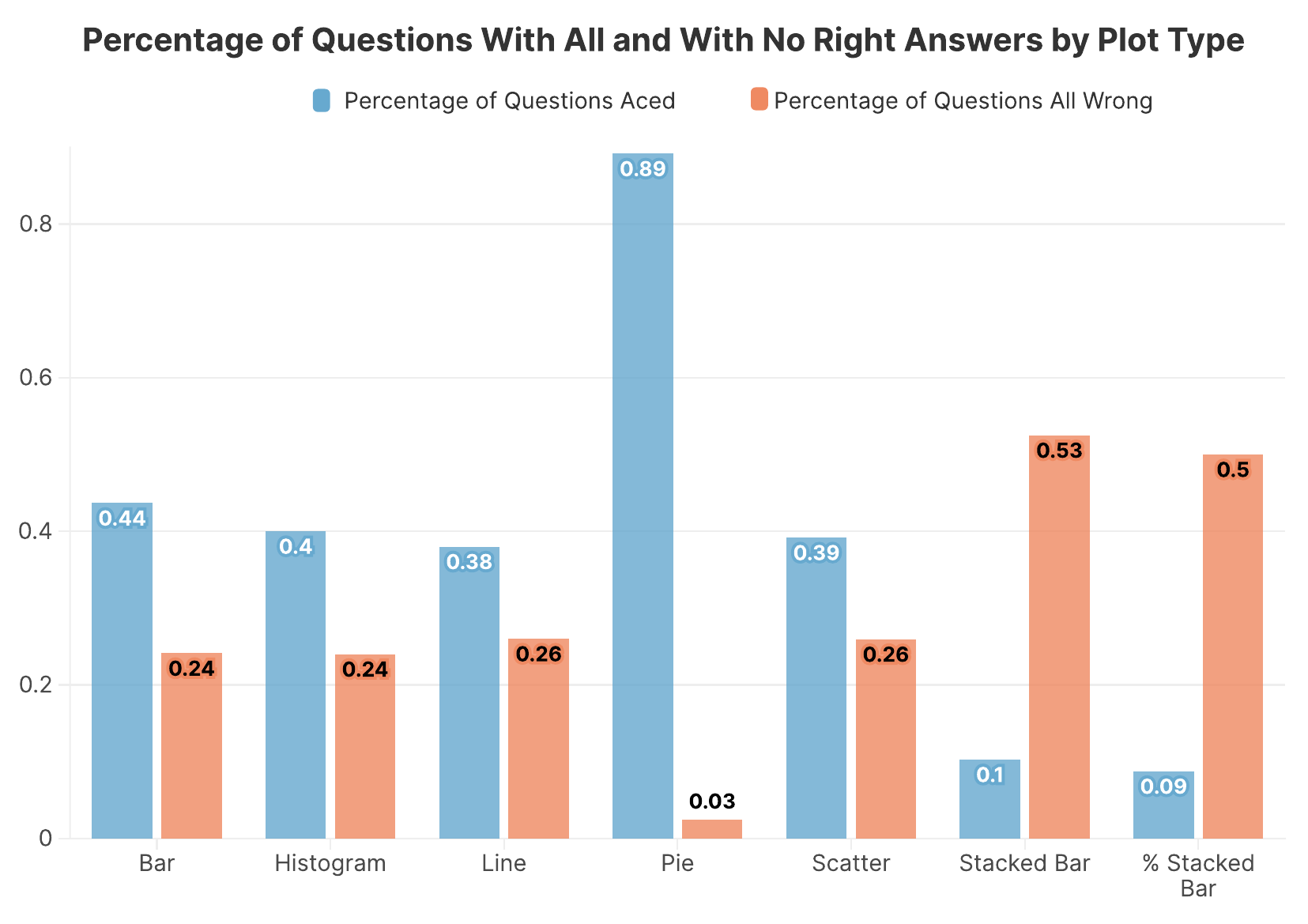} 
    \caption{\textbf{Plot type aces and all wrongs.}
    Aces and All Wrongs percentage relative to all questions, by plot types.}
    \label{fig:percplot} 
\end{figure}

\Cref{fig:percplot} further illustrates these differences by showing the percentage of questions answered correctly in all 20 iterations versus those answered incorrectly in all iterations.
\piechart led with 89\% of questions answered correctly across all attempts, while \stackedbarchart achieved only 10\% accuracy.
The omission patterns followed similar trends: highest for \stackedbarchart, lowest for \piechart, and moderate for \linechart, \barchart, and \scatterplot.

\begin{figure}[htb]
    \centering
    \includegraphics[width=0.5\textwidth]{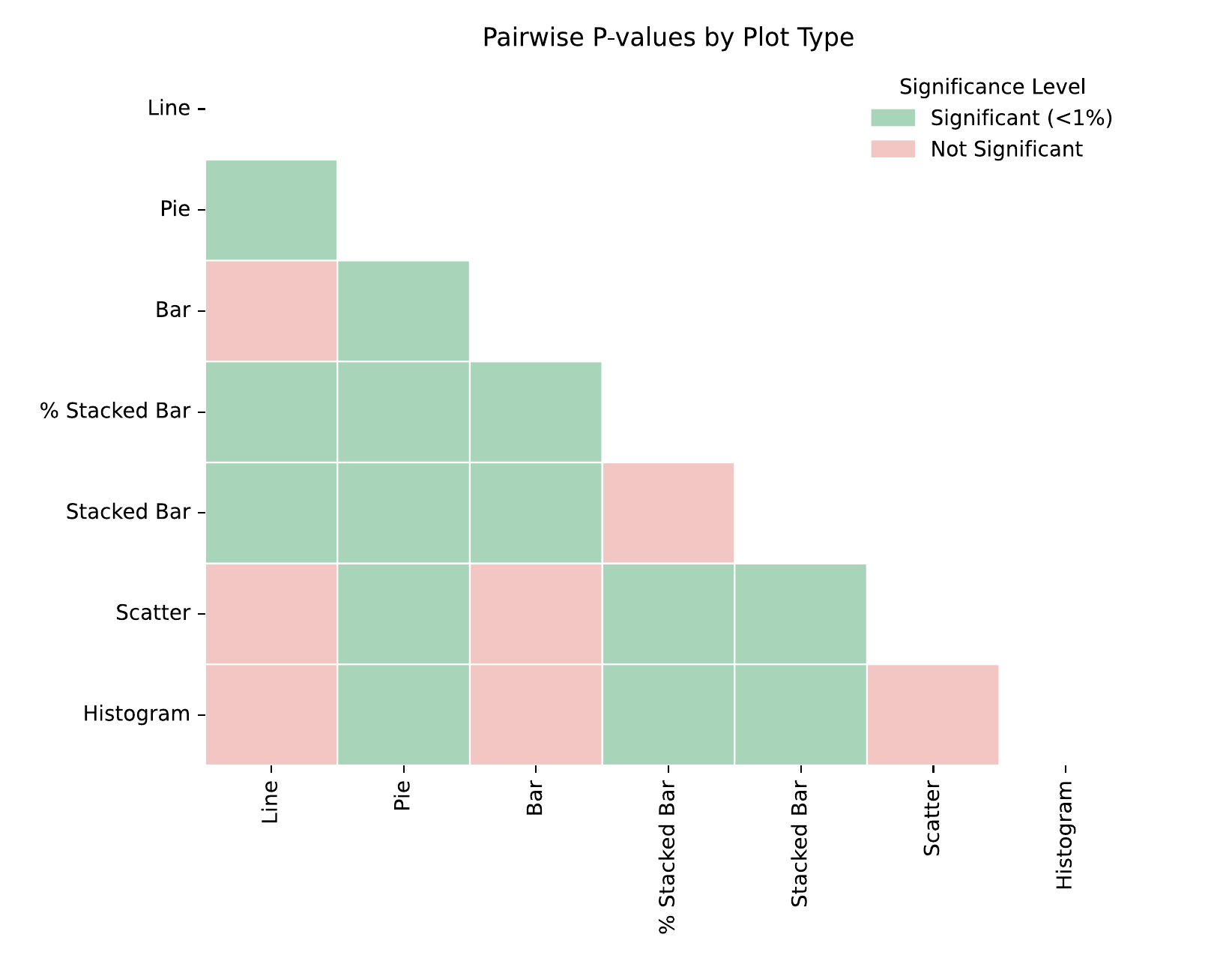} 
    \caption{\textbf{Plot types and accuracy differences.}
    Pairwise statistically different plot types.}
    \label{fig:stat_test} 
\end{figure}

We conducted statistical analyses to determine whether the models exhibited significantly different performance across plot types.
A Kruskal-Wallis test yielded a very high H-statistic, confirming that MLLMs had different accuracy medians for at least one plot type.
\Cref{fig:stat_test} presents a pairwise comparison of model performance across plot types, highlighting statistically significant differences between specific visualization formats.
These results demonstrate that plot type choice substantially influences MLLM interpretation ability, with specific formats creating unique challenges for automated analysis.

Values close to 1 (shown in blue in \cref{fig:stat_test}) indicate no statistical difference, while small values (shown in red) reveal significant statistical differences in model accuracy between plot types.
As evident in the figure, most pairings exhibit statistically different medians, confirming that MLLMs perform differently across visualization formats when other factors are controlled.
These results align with our distribution analysis: \linechart, \barchart, and \scatterplot show no significant performance differences among themselves, while \piechart produce statistically different results compared to all other formats.
The only exception involves \stackedbarchart and their percentage counterparts, which yield statistically similar performance levels---though notably, both underperform compared to other visualization types.

Building on our descriptive statistics and hypothesis testing, we conducted eight regression models using the complete dataset to investigate the specific impact of plot types while controlling for other variables. 
We focus particularly on the plot type binary markers, as they reveal the relationship between visualization format and MLLM performance when holding other factors constant.
\Cref{tab:model_results-accuracy} presents coefficients related to model accuracy, while \cref{tab:model_results-omission} shows coefficients associated with omission rates.
These regression analyses provide quantitative measurements of how each plot type influences both the correctness of MLLM responses and their tendency to abstain from answering.

\newlength{\minuslength}
\settowidth{\minuslength}{-}
\newcommand{\plus}{\hspace{\minuslength}}

\begin{table}[htb]
    \caption{\textbf{Plot type impact on MLLM accuracy.}
    Upwards arrows (↑) mean a positive significant coefficient, downwards arrows (↓) mean negative significant coefficients and sidewards arrows(→) mean non significant coefficients.
    The number of arrows indicate the size of the impact, converting to number of questions, ↑ means up to 2 more correct questions, ↑↑ means up to 4 more correct questions and ↑↑↑ means 6 or more correct questions.
    The variables under plot types are interactions between a type of question and a plot type: retrieve is short for value retrieval and compare is short for making comparisons.
    Model 1 and 3 were not controlled by datasets, Models 2 and 4 were.}
    \label{tab:model_results-accuracy}
    \centering
    \renewcommand{\arraystretch}{1.3} 
    \rowcolors{2}{White}{SteelBlue!20}    
    \resizebox{\linewidth}{!}{%
    \begin{tabular}{lcccc}
        \toprule
        \textbf{Question and Plot} & \textbf{Model 1} & \textbf{Model 2} & \textbf{Model 3} & \textbf{Model 4}\\
        \toprule
        plot\_type\_hist & \multicolumn{1}{c}{↑↑} & \multicolumn{1}{c}{↑} & \multicolumn{1}{c}{↑↑} & \multicolumn{1}{c}{↑} \\
        plot\_type\_line & \multicolumn{1}{c}{↓} & \multicolumn{1}{c}{↑} & \multicolumn{1}{c}{→} & \multicolumn{1}{c}{↑} \\
        plot\_type\_pie & \multicolumn{1}{c}{↑↑↑} & \multicolumn{1}{c}{↑} & \multicolumn{1}{c}{↑↑↑} & \multicolumn{1}{c}{↑} \\
        plot\_type\_scatter & \multicolumn{1}{c}{→} & \multicolumn{1}{c}{↑} & \multicolumn{1}{c}{→} & \multicolumn{1}{c}{↑} \\
        plot\_type\_stacked\_bar & \multicolumn{1}{c}{↓↓↓} & \multicolumn{1}{c}{↓↓} & \multicolumn{1}{c}{↓↓↓} & \multicolumn{1}{c}{↓↓} \\
        plot\_type\_stacked\_bar\_100 & \multicolumn{1}{c}{↓↓} & \multicolumn{1}{c}{↓} & \multicolumn{1}{c}{↓↓} & \multicolumn{1}{c}{↓↓} \\
        retrieve\_hist &  &  & \multicolumn{1}{c}{↑} & \multicolumn{1}{c}{↑} \\
        retrieve\_line &  &  & \multicolumn{1}{c}{↓} & \multicolumn{1}{c}{↓} \\
        retrieve\_pie &  &  & \multicolumn{1}{c}{↑} & \multicolumn{1}{c}{↑} \\
        retrieve\_scatter &  &  & \multicolumn{1}{c}{→} & \multicolumn{1}{c}{→} \\
        retrieve\_stacked\_bar\_100 &  &  & \multicolumn{1}{c}{→} & \multicolumn{1}{c}{→} \\
        compare\_hist &  &  & \multicolumn{1}{c}{↑} & \multicolumn{1}{c}{↑} \\
        compare\_line &  &  & \multicolumn{1}{c}{↓} & \multicolumn{1}{c}{↓} \\
        compare\_pie &  &  & \multicolumn{1}{c}{→} & \multicolumn{1}{c}{→} \\
        compare\_scatter &  &  & \multicolumn{1}{c}{↓} & \multicolumn{1}{c}{↓} \\
        compare\_stacked\_bar &  &  & \multicolumn{1}{c}{→} & \multicolumn{1}{c}{→} \\
        compare\_stacked\_bar\_100 &  &  & \multicolumn{1}{c}{↑↑} & \multicolumn{1}{c}{↑↑} \\
        determine\_range\_line &  &  & \multicolumn{1}{c}{↑↑} & \multicolumn{1}{c}{↑↑} \\
        determine\_range\_scatter &  &  & \multicolumn{1}{c}{↓} & \multicolumn{1}{c}{↓} \\
        \bottomrule
    \end{tabular}%
    }
\end{table}

\begin{table}[htb]
    \caption{\textbf{Plot type impact on MLLM omissions.}
    Upwards arrows (↑) mean a positive significant coefficient, downwards arrows (↓) mean negative significant coefficients and sidewards arrows(→) mean non significant coefficients.
    The number of arrows indicate the size of the impact, converting to number of questions, ↑ means up to 2 more omitted questions, ↑↑ means up to 4 more omitted questions and ↑↑↑ means 6 or more omitted questions.
    The variables under plot types are interactions between a type of question and a plot type: retrieve is short for value retrieval and compare is short for making comparisons.
    Model 1 and 3 were not controlled by datasets, Models 2 and 4 were.}
    \label{tab:model_results-omission}
    \centering
    \renewcommand{\arraystretch}{1.3} 
    \rowcolors{2}{White}{SteelBlue!20}
    \resizebox{\linewidth}{!}{%
    \begin{tabular}{p{3cm} c c c c}
        \toprule
        \textbf{Question and Plot} & \textbf{Model 1} & \textbf{Model 2} & \textbf{Model 3} & \textbf{Model 4} \\
        \toprule
        plot\_type\_hist & \multicolumn{1}{c}{↓} & \multicolumn{1}{c}{→} & \multicolumn{1}{c}{→} & \multicolumn{1}{c}{↓} \\
        plot\_type\_line & \multicolumn{1}{c}{↓} & \multicolumn{1}{c}{↑} & \multicolumn{1}{c}{↑↑} & \multicolumn{1}{c}{↓↓↓} \\
        plot\_type\_pie & \multicolumn{1}{c}{↓↓↓} & \multicolumn{1}{c}{→} & \multicolumn{1}{c}{↓↓} & \multicolumn{1}{c}{↓↓↓} \\
        plot\_type\_scatter & \multicolumn{1}{c}{↓↓} & \multicolumn{1}{c}{→} & \multicolumn{1}{c}{↑↑} & \multicolumn{1}{c}{↓↓↓} \\
        plot\_type\_stacked\_bar & \multicolumn{1}{c}{↑↑↑} & \multicolumn{1}{c}{↑↑↑} & \multicolumn{1}{c}{↑↑↑} & \multicolumn{1}{c}{↑↑↑} \\
        plot\_type\_stacked\_bar\_100 & \multicolumn{1}{c}{↑↑} & \multicolumn{1}{c}{↑↑↑} & \multicolumn{1}{c}{↑↑↑} & \multicolumn{1}{c}{↑↑} \\
        retrieve\_pie & \multicolumn{1}{c}{→} & \multicolumn{1}{c}{→} & \multicolumn{1}{c}{↓↓} & \multicolumn{1}{c}{↓} \\
        retrieve\_scatter & \multicolumn{1}{c}{→} & \multicolumn{1}{c}{→} & \multicolumn{1}{c}{↓↓↓} & \multicolumn{1}{c}{↓↓↓} \\
        retrieve\_stacked\_bar\_100 & \multicolumn{1}{c}{→} & \multicolumn{1}{c}{→} & \multicolumn{1}{c}{↑↑↑} & \multicolumn{1}{c}{↑↑↑} \\
        compare\_hist & \multicolumn{1}{c}{→} & \multicolumn{1}{c}{→} & \multicolumn{1}{c}{↑} & \multicolumn{1}{c}{↑} \\
        compare\_line & \multicolumn{1}{c}{→} & \multicolumn{1}{c}{→} & \multicolumn{1}{c}{↓↓} & \multicolumn{1}{c}{↑} \\
        compare\_pie & \multicolumn{1}{c}{→} & \multicolumn{1}{c}{→} & \multicolumn{1}{c}{↓↓↓} & \multicolumn{1}{c}{↓↓} \\
        compare\_scatter & \multicolumn{1}{c}{→} & \multicolumn{1}{c}{→} & \multicolumn{1}{c}{↓↓} & \multicolumn{1}{c}{↑↑↑} \\
        compare\_stacked\_bar & \multicolumn{1}{c}{→} & \multicolumn{1}{c}{→} & \multicolumn{1}{c}{↑↑↑} & \multicolumn{1}{c}{↑↑} \\
        compare\_stacked\_bar\_100 & \multicolumn{1}{c}{→} & \multicolumn{1}{c}{→} & \multicolumn{1}{c}{↓↓↓} & \multicolumn{1}{c}{↓↓↓} \\
        determine\_range\_line & \multicolumn{1}{c}{→} & \multicolumn{1}{c}{→} & \multicolumn{1}{c}{↓↓} & \multicolumn{1}{c}{↑} \\
        \bottomrule
    \end{tabular}%
    }
\end{table}

This being an effects-coded regression, binary coefficient values, such as plot type binaries and their interactions, are to be interpreted as deviations from the model's grand mean---the overall mean across all observations (size-weighted due to unequal subgroup sizes).

The regression results clearly show that plot types significantly impact both MLLM accuracy and omission patterns.
In seven of our eight models, at least four different plot types showed statistically significant non-zero coefficients, and all models had at least two significant plot type effects.
The magnitude of these effects varied substantially.
Omissions' Model 2 shows that \percbarchart increases omission by 1.17 standard deviations compared to the mean---translating to approximately 10 more omitted answers out of 20 attempts (given the standard deviation of 8.5).
In contrast, Model 3 indicates only a minor increase of 0.16 standard deviations (roughly one question) in accuracy when analyzing \linechart versus the model mean.

The plot type variables reveal consistent patterns on accuracy; most of their statistically significant coefficients maintained the same sign through different regressions.
Out of the 6 plot type binaries, only \linechart presented both positive and negative coefficients.
Even then, \linechart impact was consistent: having a negative impact on MLLM accuracy in both models without dataset control, and a significant positive impact in the controlled specifications.

For accuracy, \stackedbarchart and \percbarchart lead MLLMs to worse outcomes than the grand mean, while \histogram and \piechart boosted MLLMs correctness the most, yielding above mean performance in all regression specifications. 

In terms of omissions, plot types were slightly less consistent, with \linechart and \scatterplot displaying positive and negative results along specifications.
All others maintained the same sign during control variation, with \stackedbarchart and \percbarchart once again being a negative highlight.
The regressions show their significant role in driving MLLM omissions above the grand mean, with \stackedbarchart having all coefficients above 0.68 standard deviations.
\piechart had the most beneficial impact on omissions: MLLMs omit around 0.3 standard deviations less using them. 

The interaction variables also reveal interesting patterns, with 8 of 13 interaction terms achieving statistical significance in both accuracy models.
Similarly, in omission models, 6 of 10 and 8 of 10 interactions were significant, mostly with negative coefficients.

Among the accuracy interactions, the most striking finding involves the comparison questions combined with \percbarchart.
Despite its negative main effects commented above, it demonstrated a substantial positive interaction effect (1.32 standard deviations) for comparison questions.
This finding aligns with the significant negative coefficient for this combination in the omission models, suggesting MLLMs are both more accurate and more confident when evaluating comparisons using percentage-based visualizations.

Comparison questions with \stackedbarchart lead to an increase in omissions, despite the plot similarities with \percbarchart, and no combinations increased accuracy while reducing omissions.
Many had beneficial impacts on MLLMs performance, be it through reducing omissions (retrieval questions in \scatterplot) or increasing accuracy (retrieval questions with \piechart and with \histogram).
\scatterplot was the only plot type to have interactions that shift given model specifications (with comparison questions in the omissions regressions) and the only model to have negative results on both accuracy and omissions (via Model 3).

Despite the inconsistency of two plot type coefficients, and one interaction, our model specifications showed robustness.
The inclusion of interaction terms (Models 3 and 4) preserved the direction and significance of most coefficients (altering 0 out of 6 in accuracy and 2 out of 6 in omissions) from the simpler models.
Similarly, controlling for dataset characteristics (Models 2 and 4) did not substantially alter the coefficients (altering 2 out of 19 for accuracy and 1 out of 16 for omissions), demonstrating consistency 
independent of data contexts.

\subsection{Impact of Color Palette}

\begin{figure}[htb]
    \centering
    \includegraphics[width=0.5\textwidth]{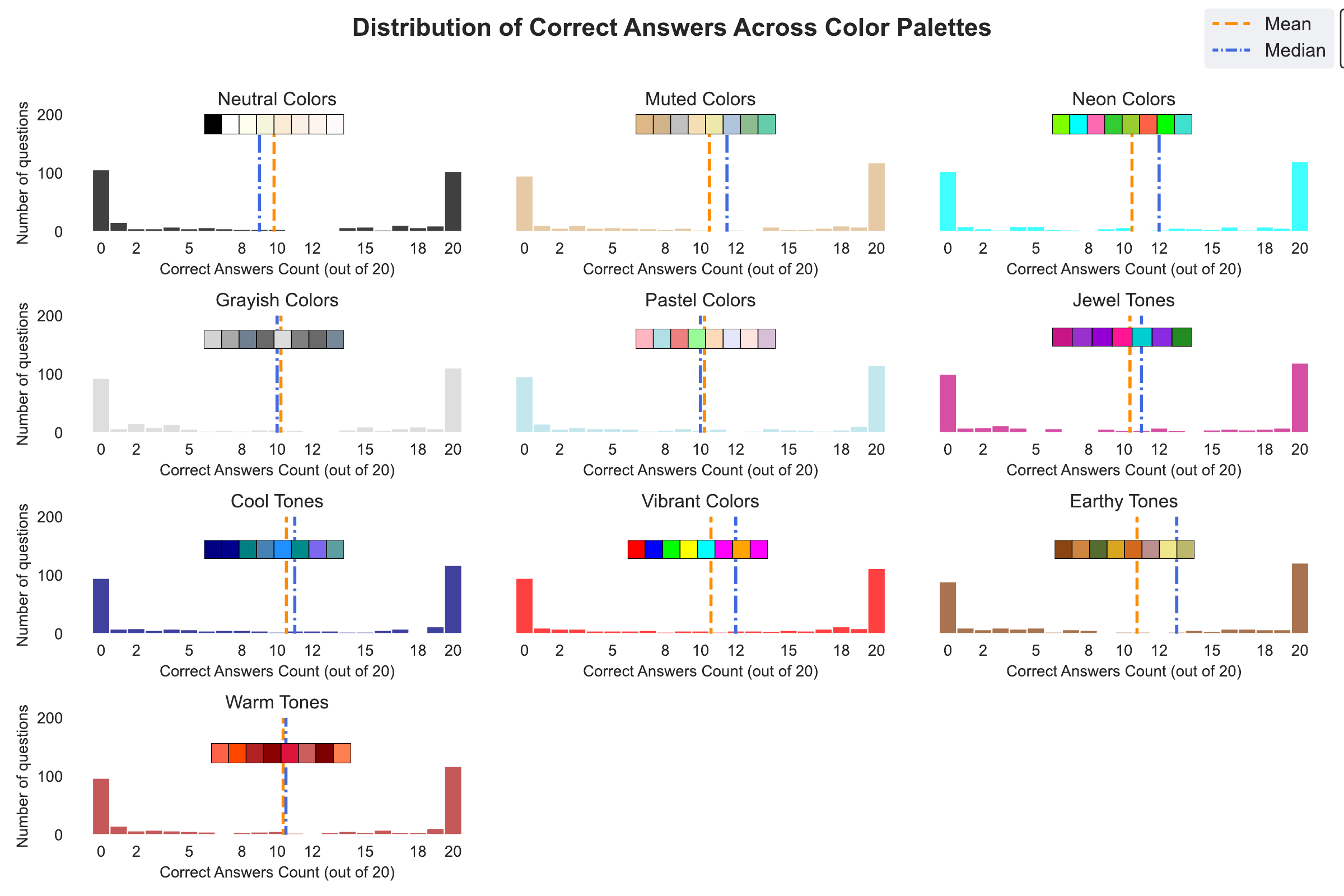} 
    \caption{\textbf{Accuracy over color.}
    Performance for different color palettes.}
    \label{fig:hist_colors}
\end{figure}

To investigate how \colorpalette affect MLLM performance, we analyzed accuracy and omission statistics when models interpreted visualizations with different color schemes.
\Cref{fig:hist_colors} displays the distribution of questions with varying accuracy levels across different \colorpalette{s}.
The distributions appear remarkably similar across all color schemes, suggesting minimal influence of color on MLLM interpretation abilities. 
Statistical analysis confirms this observation: average accuracy ranged narrowly from 10.0 to 10.9 correct answers across all \colorpalette{s}.
Median values showed slightly more variation, from 10 correct answers for black palettes to 14 for saddlebrown palettes, but these differences lack statistical significance.

This consistency in performance extends to omission patterns.
Models showed similar abstention tendencies regardless of \colorpalette, with only minor variations in average and median omission rates.
This suggests that, unlike plot type, there is little evidence that \colorpalette choice has an impact on MLLM visualization literacy.

To confirm these observations, we conducted statistical tests examining whether MLLMs showed significantly different performance across \colorpalette.
The Kruskal-Wallis test yielded a high p-value of 0.97 and a small H-statistic of 2.84 for accuracy---directly contrasting with our plot type findings and strongly indicating no statistical difference in MLLM accuracy across \colorpalette{s}.
For omissions, we observed a slightly higher H-statistic of 6.54, but the p-value remained high at 0.68, confirming that omission patterns also show no significant variation across color schemes.
Our regression analysis further reinforced this conclusion.
Across all 14 model specifications we tested, not a single \colorpalette binary variable achieved statistical significance.

\subsection{Impact of Title}

Next, we examine how changing plot \titlefactor from neutral descriptions to suggestive ones (which hint at findings in the visualization) affects MLLM performance.
We follow the same analytical structure as before.

\begin{figure}[htb]
    \centering
    \includegraphics[width=0.5\textwidth]{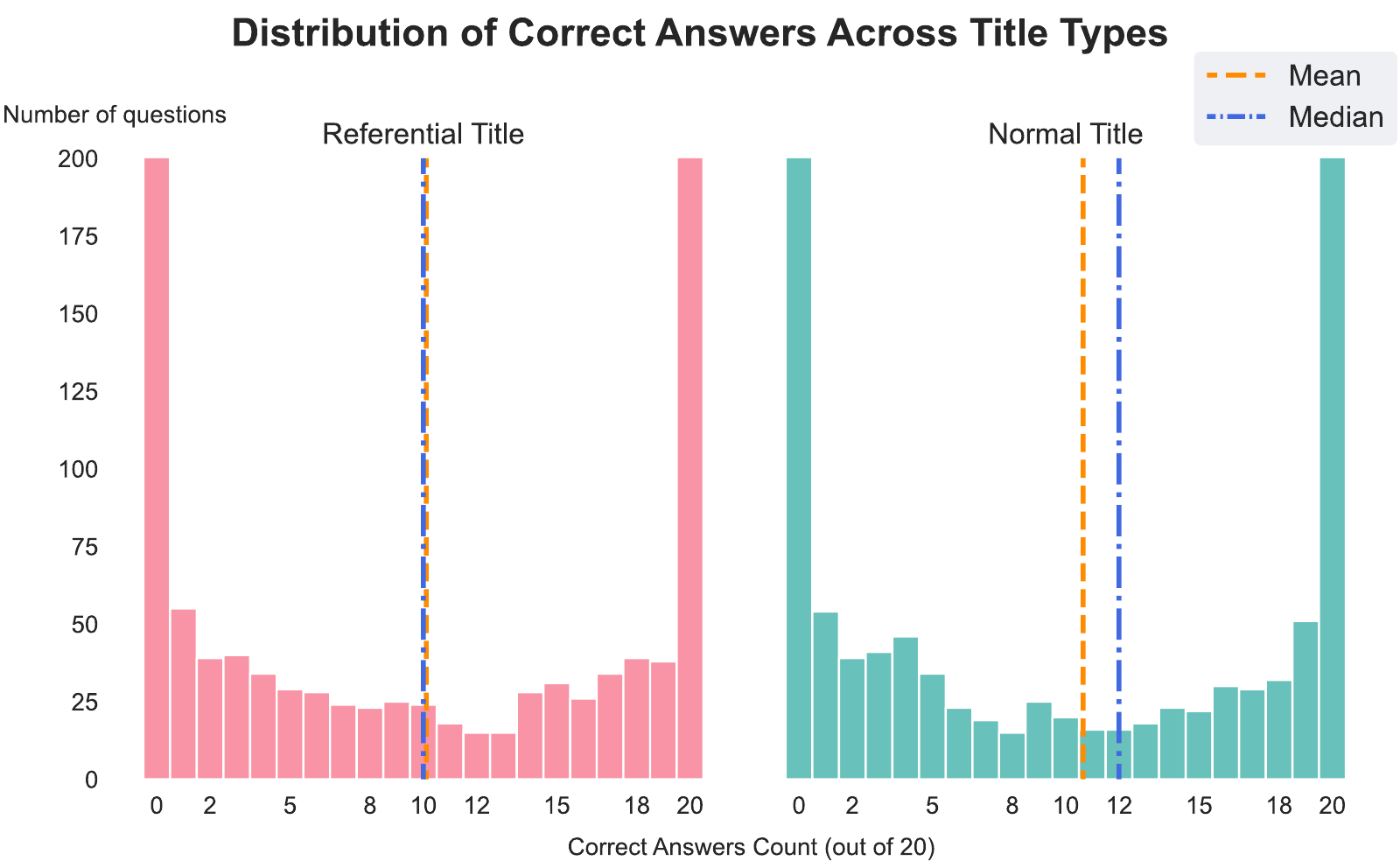} 
    \caption{\textbf{Plot title impact.}
    Accuracy distribution with different title types.}
    \label{fig:histTitle}
\end{figure}

\Cref{fig:histTitle} shows the distribution of questions by number of correct answers, aggregated by \titlefactor type.
The distributions for normal and suggestive titles share many similarities.
Their central tendencies are quite close, though normal titles show a slight performance advantage with a median of 12 correct answers compared to 10 for suggestive titles.
Both distributions exhibit a bimodal shape with peaks at 20 and 0 correct answers.
Unlike the plot type and color palette histograms, these distributions feature a notably higher percentage of questions answered completely incorrectly, suggesting that title variations may influence model performance differently from other visualization attributes.

These apparent similarities prompted us to conduct statistical testing to determine whether MLLMs demonstrate significantly different performance when interpreting plots with suggestive titles.
For accuracy, the Kruskal-Wallis test yielded an H-statistic of 3.834 and a p-value of exactly 0.05, indicating a borderline statistically significant difference.
This threshold p-value suggests that MLLMs may perform differently with suggestive titles, though the evidence is not definitive.
For omission rates, we found a lower H-statistic of 1.457 and a p-value of 0.227, more clearly indicating that MLLMs do not exhibit significantly different omission patterns when presented with suggestive titles.

Given these ambiguous initial results, we turned to regression analysis for more nuanced insights.
Across all eight models, suggestive titles consistently showed statistically significant coefficients---negative for accuracy and positive for omissions.
Remarkably, these coefficients were highly consistent: -0.06 standard deviations for accuracy models and 0.07 standard deviations for omission models, regardless of other model specifications.
These consistent findings provide stronger evidence that suggestive titles have a small but reliable negative impact on MLLM performance, leading to slightly reduced accuracy and slightly increased tendency to omit responses when interpreting visualizations.

\subsection{Impact of Within-Task Plot Type Substitution}

Not all plot types in our dataset are logically interchangeable due to dimensionality constraints and underlying data structures.
To investigate the effects of plot type substitution more precisely, we grouped visualizations into categories where comparisons can be made:

\begin{enumerate}[noitemsep, topsep=0pt, parsep=0pt, partopsep=0pt]
    \item Categorical (\piechart, \barchart, \scatterplot);
    \item Unidimensional  (\linechart, \barchart, \scatterplot); and
    \item Multidimensional (\stackedbarchart, \percbarchart, \linechart).
\end{enumerate}

Within each group, we conducted dummy-coded regression analyses to assess how MLLM performance varies across plot types, using a common visualization format as the reference category.
This approach ensures fair comparisons when evaluating how plot type selection impacts MLLM interpretation capabilities for the same underlying data.
The results are presented in table \ref{tab:final_compact}.

\begin{table}[htb]
    \caption{\textbf{Plot by plot standoff.}
    Arrows that are not sideways indicate significance, upwards (↑) means positively significant, downwards (↓) means negatively significant.
    If the effect absolute value is less than 0.3, one arrow (↑), between 0.3 and 0.6 two (↑↑), and more three arrows (↑↑↑).
    ``omi'' is omission, ``acc'' accuracy, Cat.\ Multi.\ and Uni.\ refer to the visualization groups.}
    \label{tab:final_compact}
    \centering
    \scriptsize
    \begin{tabular}{lcccccc}
        \toprule
        & \multicolumn{2}{c}{\textbf{Cat.}} & \multicolumn{2}{c}{\textbf{Uni.}} & \multicolumn{2}{c}{\textbf{Multi.}} \\
        \cmidrule(lr){2-3} \cmidrule(lr){4-5} \cmidrule(lr){6-7}
        \textbf{Plot} & acc & omi & acc & omi & acc & omi \\
        \midrule
        \piechart & → & → & -- & -- & -- & -- \\
        \scatterplot & ↓ & → & → & → & -- & -- \\
        \barchart & -- & -- & → & → & -- & -- \\
        \linechart & -- & -- & -- & -- & ↑↑↑ & ↓ \\
        \percbarchart & -- & -- & -- & -- & ↑ & ↓↓↓ \\ \\
        \bottomrule
    \end{tabular}
\end{table}

\vspace{-0.1cm}

MLLMs demonstrate clear preferences for certain plot types to achieve higher accuracy in specific tasks.
\Cref{tab:final_compact} shows us that MLLMs have a small, but statistically significant, preference for \barchart and \piechart over \scatterplot when handling displays of  categorical elements. 
We also see that MLLMs are indifferent between \barchart, \scatterplot and \linechart when going over uni-dimensional time series, and that when handling multiple dimensions over time, the models have significantly better performances with \linechart (0.65 standard deviations better) and \percbarchart (0.17 standard deviations better) than with \stackedbarchart.

MLLMs exhibit a different behavior towards omission (\Cref{tab:final_compact}).
On average, and holding other factors constant, they omit the same in \barchart, \piechart and \scatterplot when handling categorical visualizations.
The same indifference arises in the interpretation of time-series visualizations: MLLMs do not have a significant difference in omission whether the plot is a \linechart, a \barchart or a \scatterplot. 
Lastly, they do maintain strong preferences over multidimensional visualizations: being presented with a \linechart decreases the omission in incredible 0.97 standard deviations when compared to a \stackedbarchart.
Considering an 8.5 standard deviation, that's roughly 8 less omissions per question. \percbarchart are also a big improvement over their non-percentual counterpart: MLLMs omit 0.27 standard deviations less when using \percbarchart.
Such impact also strengthens the findings \linechart{s} and \percbarchart{s} had on accuracy.

\section{Discussion}

Our statistical analysis reveals several patterns in how visualization characteristics affect MLLM performance.
We found that chart type significantly influences both accuracy and omission rates, with MLLMs demonstrating higher accuracy on simpler visualizations like \piechart{s} while struggling with more complex types such as \stackedbarchart{s}.
Interestingly, specific combinations of chart types and analytical tasks showed distinctive performance---for instance, \percbarchart yielded strong results for comparison.

Two particularly noteworthy findings emerged from our analysis.
First, \colorpalette{s} showed no significant impact on MLLM performance across all tested conditions. 
Second, suggestive \titlefactor{s} consistently increased omission rates and reduced accuracy across models, suggesting that MLLMs may become more hesitant when presented with potentially misleading contextual information.

\subsection{Why Changing Plot Type Affects MLLM Performance}

Different possibilities explain why plot types significantly influence MLLM performance.
We first speculate that \textbf{the visual representation of the plot substantially impacts the image vectorization process}.
This could explain why more subtle visual changes---like color palette variations---don't significantly affect model performance.
A potential counterpoint is that titles, despite their minimal visual footprint in the image, showed high significance for both accuracy and omissions.
However, this counterpoint isn't conclusive since natural language likely holds substantial weight in the vectorization process.
To further validate this speculation, we would need to investigate which elements most significantly affect the vectorization process and by what magnitude.

Our second speculation attributes \textbf{plot type effects to model training and chart prevalence on the web}. 
Plot types more commonly found across internet resources and training corpora likely present fewer interpretation challenges for MLLMs.
This explanation aligns with our observation that less common visualizations like \stackedbarchart{s} and \percbarchart{s} yielded worse results, while ubiquitous formats like \piechart{s} led to better performance. It is also supported by the very VLAT paper~\cite{lee_vlat_2017}, where they justify the inclusion of the different visualization types by general popularity and usage, in their rankings \piechart{s} rank higher along \barchart and \linechart, while \stackedbarchart and \percbarchart are above \scatterplot and \histogram, but in the lower half.

Our third speculation suggests that plot types themselves might not be the decisive factor in MLLM performance---the \textbf{underlying dataset complexity could be the true differentiator}.
Some chart types consistently associated with poor performance, such as \stackedbarchart{s}, typically represent more complex, higher-dimensional data.
Two findings support this hypothesis: first, all dataset control variables showed highly significant coefficients in our regression models; second, including these dataset controls often lowered the plot types' coefficients significancy, suggesting the primary effect being captured was the dataset (and its complexity) rather than the visualization type. 
This possibility also raises the discussion of whether test sets that vary their underlying data to such different degrees of complexity should be testbeds for MLLM visualization literacy.

\subsection{The Title Effect}

Title framing significantly influencing how MLLMs interpret charts aligns with human visualization literacy research, where Kong et al.~\cite{kong18} demonstrated that title framing can substantially shape the message a person perceives from a chart.
Even more compelling, when a title contradicts the actual visual content, human recall tends to align more closely with the title than with the data representation itself~\cite{kong2019misalignment}.

Our results suggest that suggestive titles increase model uncertainty, leading to higher omission rates. 
This indicates that MLLMs may detect conflicts between textual framing and visual data, triggering a more cautious response pattern---similar to how humans might question contradictory information.
This finding has implications for visualization design in contexts where automated interpretation is expected.

\subsection{The Color Palette (Lack of) Effect}

Our experiments provide overwhelming evidence that color palette variations do not significantly affect MLLM accuracy or omission rates when answering visualization questions.
This finding stands in stark contrast to human visualization literacy research, which demonstrates that certain color choices---particularly bright or low-contrast palettes---can significantly impair human chart comprehension~\cite{Rhyne2016}.

Within MLLM research, Li et al.~\cite{Li2024} hypothesized that color similarity might cause models to confuse different categories in \stackedbarchart{s}.
Bendeck and Stasko ~\cite{DBLP:journals/tvcg/BendeckS25} also state that MLLMs struggle with color differentiation.
Our findings strongly contradict those hypothesis, suggesting that color palette choices are largely irrelevant to MLLM performance.
We propose two potential explanations:

\begin{enumerate}[noitemsep, topsep=0pt, parsep=0pt, partopsep=0pt]
    \item \textbf{Vectorization mechanics:} Colors may not substantially alter the structural properties of the input vector compared to other graphical elements.
    While plot type transforms the essential structure of the visual encoding, color changes represent more superficial variations that preserve the underlying spatial relationships.

    \item \textbf{Training exposure:} MLLMs likely encountered numerous charts with diverse color schemes during training, potentially learning to focus on structural patterns rather than chromatic attributes.
\end{enumerate}

An additional factor may be the nature of the VLAT dataset itself.
No plot in our study relies heavily on color-specific semantics (e.g., there are no cases where green specifically signals approval while red signals rejection).
As long as the color palette establishes visual differentiation between elements, the specific colors used may not matter.
Moreover, several plot types in our study---including \linechart, \barchart, \histogram, and \scatterplot---do not primarily use color to encode meaning, limiting the potential for color palette effects to manifest.

\subsection{Save an MLLM, Change a Plot}

Our findings provide clear guidance for optimizing visualizations for MLLM interpretation: 
the most important factor is to choose plot type wisely.
Our results consistently demonstrate that while colors have little impact on MLLM performance and titles have significant but small effects, plot type substantially influences both accuracy and omission.

The plot thickens, however, when determining the optimal plot type.
We observed several context-dependent exceptions to general patterns:

\begin{itemize}[noitemsep, topsep=0pt, parsep=0pt, partopsep=0pt]
    \item \percbarchart outperformed \scatterplot and \linechart for comparison tasks; and
    \item In subset analyses, \piechart did not consistently outperform \barchart as dramatically as expected.
\end{itemize}

Despite these edge cases, certain plot types demonstrated consistently superior performance across questions, datasets, and variations:

\begin{itemize}[noitemsep, topsep=0pt, parsep=0pt, partopsep=0pt]
    \item \piechart{s} yielded good results across many conditions;
    \item \histogram{s} improved accuracy while reducing omissions; and
    \item \percbarchart{s} consistently outperformed regular \stackedbarchart{s}.
\end{itemize}

Our work confidently recommends specific transitions for certain data types: \linechart{s} for multidimensional time-series, and \barchart{s} or \piechart{s} for categorical datasets.
However, due to plot interchangeability limitations in our experimental design, we cannot definitively claim that high-performing plots would maintain their advantage across all possible visualization scenarios.
However, the consistency of our results strongly suggests that the performance patterns identified would likely persist across broader contexts.

\subsection{Limitations}

Our work faces a fundamental tension between methodological rigor and experimental scope.
On one hand, using the established VLAT test set provides a validated framework for assessing visualization literacy, enabling direct comparisons with human performance benchmarks.
On the other hand, our need for robust statistical analysis required generating more data points through systematic variations.

Despite our intention to isolate the specific effect of each plot type while holding all other variables constant, the inherent structure of the VLAT dataset imposed constraints on our experimental design.
Not every visualization could be transformed into every plot type due to dimensionality requirements of the underlying data.
Because we prioritized using verified, meaningful questions that genuinely assess visualization literacy, we could not extend the VLAT test set arbitrarily.

This limitation affects some aspects of our regression analysis, where plot type effects may incorporate spillover effects from other variables, particularly dataset characteristics.
While our methodology remains statistically sound, this constraint necessitates caution when interpreting the precise magnitude of certain effects.

\section{Conclusion and Future Work}

Our systematic investigation of visualization characteristics and their influence on MLLMs offers several important insights for both AI development and visualization design.
We found that plot type significantly impacts MLLM performance, with \piechart{s} yielding the highest accuracy and \stackedbarchart{s} proving to be the most challenging.
Interestingly, \colorpalette{s} showed no significant effect on interpretation capabilities of MLLMs, contrary to our initial expectations.
Meanwhile, the type of \titlefactor does not affect the overall accuracy, but does influence the tendency of the model to omit responses when faced with suggestive framing.
These findings point toward a convergence between human and machine visualization literacy, suggesting that visualizations designed with established human perceptual principles often work well for MLLMs too---the plot is thickening, indeed.


In our future work, we plan to build upon our findings by exploring several promising directions.
Firstly, investigating more complex visualization types beyond the standard VLAT set would extend our understanding to more sophisticated data representations, including 3D and immersive analytics charts as well as novel visualization types, especially those increasingly used in data journalism.
Secondly, evaluating how MLLMs perform on visualizations with deliberate errors or misleading elements could advance our knowledge about their robustness and susceptibility to visual deception.
Additionally, directly comparing human and MLLM performance on identical visualization tasks would illuminate differences in perception strategies and error patterns.
And what about interactive visualizations?
It would be worth building on the new trend of agentic AI in investigating how an MLLM could autonomously interact with a visualization.
Finally, developing specialized training techniques---essentially prompt engineering---to improve MLLM interpretation of challenging chart types could lead to more consistent performance across visualization formats.



\acknowledgments{%
    This work was supported by Villum Investigator grant VL-54492 by Villum Fonden.
    Any opinions, findings, and conclusions expressed in this material are those of the authors and do not necessarily reflect the views of the funding agency.
    The first author would like to thank Zofia Szulc, Niklas Elmqvist, Ogum and all others who allowed him to stay in Brazil during the writing of this paper and recover his mental health.
}

\bibliographystyle{abbrv-doi-hyperref}
\bibliography{plot-thickens}

\end{document}